\documentclass[10pt,conference]{IEEEtran}
\usepackage{amsmath,amssymb}
\usepackage{bbm}
\usepackage{graphicx}
\usepackage[noadjust]{cite}
\usepackage{subfig}
 
\usepackage{epstopdf}
\usepackage{float}
\usepackage{mathrsfs}
\usepackage{psfrag}
\usepackage[below]{placeins}
\newcommand{\peqnap}{\overline{P}_{A}}
\newcommand{\peqnsta}{\overline{P}_{S}}
\newcommand{\pcondeqnap}{\overline{P}_{AP}^{(s_i; l_1,l_2, l_3)}}
\newcommand{\pcondeqnsta}{\overline{P}_{STA}^{(s_i; l_1,l_2, l_3)}}
\newcommand{\papwinms}{\hat{P}_{AP, MS}^{(s_i; l_1,l_2, l_3)}}
\newcommand{\pstawinms}{\hat{P}_{STA, MS}^{(s_i; l_1,l_2, l_3)}}
\newcommand{\pcondeqnapend}{\overline{P}_{AP}^{(s_j; l_1-a+c,l_2-b+d, l_3+a+b+e)}}

\newcommand{\pap}{P_A}
\newcommand{\psta}{P_S}

\newcommand{\iid}{{i.~i.~d.}}
\newcommand{\EX}{\mathbb{E}}

\newlength{\eqboxstorage}

\newlength{\eqboxstoragel}

\newcommand{\pcol}{P_{col}}
\newcommand{\tcol}{T_{col}}

\newcommand{\perr}{P_{err}}

\newcommand{\wpr}{\it w.p.}
\newcommand{\renew}{{R}}
\newcommand{\state}{\mathcal{S}}
\newcommand{\idletime}{\mathcal{I}}
\newcommand{\psucap}{P_{suc, AP}^{i}}
\newcommand{\psucsta}{P_{suc,STA}^{i}}

\newcommand{\tsuc}{T_{suc}}

\IEEEoverridecommandlockouts

\begin{document}
\title{A Joint Uplink/Downlink Opportunistic Scheduling Scheme for Infrastructure WLANs}
\author{ 
\IEEEauthorblockN{Nischal S and Vinod Sharma}
\IEEEauthorblockA{Department of Electrical Communication Engineering,\\
Indian Institute of Science, Bangalore 560012, India.\\
Email: nischal, vinod@ece.iisc.ernet.in}

}
\maketitle
\begin{abstract}
We propose a combined uplink/downlink opportunistic scheduling algorithm for infrastructure WLANs.
In the presence of both uplink and downlink flows, an infrastructure WLAN suffers from the uplink/downlink unfairness problem which severely decreases the throughput of the access point (AP).
We resolve the unfairness by maintaining a separate queue and a backoff timer for each associated mobile station (STA) at the AP.
We also increase the system throughput by making the backoff time a function of the channel gains. This reduces the collision probability also.
We theoretically analyze the performance of the system under symmetric statistics for all users and validate the analysis by extensive simulations.
Simulation results show increase in system throughput by over $40\%$ compared to the 802.11 MAC.

\end{abstract}

\begin{keywords}
 WLAN, Opportunistic scheduling.
\end{keywords}

\section{Introduction}
\label{sec:Intro}
Wireless local area networks (WLANs) based on the IEEE 802.11 family of standards have seen rapid proliferation in the past few years. From the time the original version of the standard IEEE 802.11 was released in 1997, 
IEEE 802.11 family of standards has been continuously evolving with the release of several substandards and amendments to meet the evergrowing traffic demands.
Performance of DCF, the default MAC protocol in infrastructure WLANs has been the subject of numerous research studies (\cite{bianchi00,kumar05,kuriakose07}). 
The DCF protocol is well-known to be inefficient and to cause significant fairness issues in the infrastructure WLANS(\cite{Heusse03,kim2005downlink,nischal13}).
 In this paper we address both of these issues.
First we review the relevant literature.

\subsection{Opportunistic scheduling}
When there are multiple nodes contending for access to a channel, a subset of these nodes will see a relatively good channel quality to their destination. 
Opportunistic scheduling algorithms aim to take advantage of the instantaneous channel variations by scheduling the users with good channels to their destination for transmission.
Nodes with good channel conditions can transmit at higher data rates for a given error performance, thereby maximising the system throughput.

A multitude of opportunistic scheduling schemes have been proposed for IEEE 802.11 WLANs in the literature.
Most of these schemes, e.g., MAD\cite{Ji04}, OSMA\cite{wangOSMA}, WDOS\cite{DBLP:conf/ifip6-8/HahmLK06} and CFCA\cite{Kim06:37} perform opportunistic scheduling only for the downlink.
Some opportunistic schemes, e.g. O-CSMA/CA\cite{cioffiuplink} have been proposed exclusively for the uplink.
To the best of our knowledge, only \cite{Yoo08} proposes opportunistic scheduling for both the uplink and the downlink.

   The authors in \cite{Yoo08} argue that, in \emph{downlink-only} opportunistic scheduling schemes, without proper scheduling in the uplink, the gain from opportunistic scheduling reduces as the number of stations increase. 
This is because the AP gets fewer chances to transmit its packets due to the uplink-downlink unfairness in the infrastructure WLAN (section \ref{sec:ul_dl}). 

Another challenge in designing opportunistic scheduling for WLANs is obtaining timely channel state information (CSI) at the transmitters.
The opportunistic scheduling schemes mentioned above differ in the method by which the CSI is acquired.
Early opportunistic scheduling schemes like MAD\cite{Ji04}, OSMA\cite{wangOSMA}, WDOS\cite{DBLP:conf/ifip6-8/HahmLK06} send a multicast RTS (MRTS)
containing a prioritized list of potential receivers.
These receivers then estimate their channel quality from the MRTS packet and reply back with CTS to the AP in an order based on their priority and/or the channel gain from the AP.
Such schemes incur large overheads due to channel probing and feedback which greatly reduces the gains obtained due to opportunistic scheduling.
Moreover, such schemes cannot be implemented in the uplink.
 
   On the other hand, in CFCA\cite{Kim06:37}  the channel gain is estimated based on the SNR of the CTS and ACK control packets.
The authors in \cite{Kim06:37} claim that the error in channel estimation is within the acceptable range.
Furthermore, \cite{DBLP:conf/vtc/DianatiT08} shows that opportunistic scheduling using CSI obtained using exisiting ARQ signals like ACK achieve significant throughput gains by 
exploiting a reasonable level of multiuser diversity without the complexity of having explicit feedback. 

\subsection{Uplink Downlink Unfairness}
\label{sec:ul_dl}
A serious unfairness betweeen uplink and downlink flows is seen in infrastructure WLANs with IEEE 802.11 DCF MAC scheme.
In an infrastructure WLAN, the access point (AP) carries traffic for all the mobile stations (STAs) in the downlink. 
It competes with STAs in the uplink using the DCF MAC protocol to gain access to the channel. 
However, the DCF protocol ensures equal probability of channel access to all competing stations irrespective of their offered load. In other words, the long term probability that the AP wins contention is $\frac{1}{N+1}$,
where  $N$ is the number of mobile stations eventhough it may have much more traffic to transmit than the STAs.  

A number of papers have addressed the issue of uplink/downlink unfairness in 802.11 WLANs at the MAC layer (\cite{lopez2008,hirantha2008dynamic,kim2005downlink,gopalakrishnan2004,keceliweighted}) 
as well as at the upper layers (\cite{pilosof2003understanding,wu2005upstream,ha2006wlc29}).

The MAC layer solutions to the unfairness problem usually either assign higher priority to the AP by proper tuning of the backoff parameters or the interframe spaces (\cite{lopez2008,hirantha2008dynamic,kim2005downlink}) or 
allow the AP to send more data packets when it gets the access to the channel (\cite{gopalakrishnan2004,keceliweighted,hiraguri2013}). 

In \cite{siwam2008}, the authors maintain a separate queue at the AP and a separate random backoff timer for each mobile station.
The advantage of such a scheme is that it requires minimal changes over the existing 802.11 DCF scheme.
However it also increases the probability of collision. 
We use a similar method wherein the AP maintains a separate queue and backoff timer for each mobile station. 
However our timer is not random  and is opportunistic thereby reducing collisions and increasing throughput. 

The rest of the paper is organized as follows. The system
and the channel model is explained in Section \ref{sec:sysmodel}. 
The SNR to timer mapping used in our opportunistic scheduling scheme is described in Section \ref{sec:timerfunc}. 
A detailed performance analysis of the system is carried out
in Section \ref{sec:fixedpoint}. The results of the simulations are provided in
Section \ref{sec:oppsimln}. Finally, conclusions are drawn in Section \ref{sec:conclusion}.

\section{System Model}
\label{sec:sysmodel}

\begin{figure}[!h]
 \centering
 \includegraphics[scale=0.6]{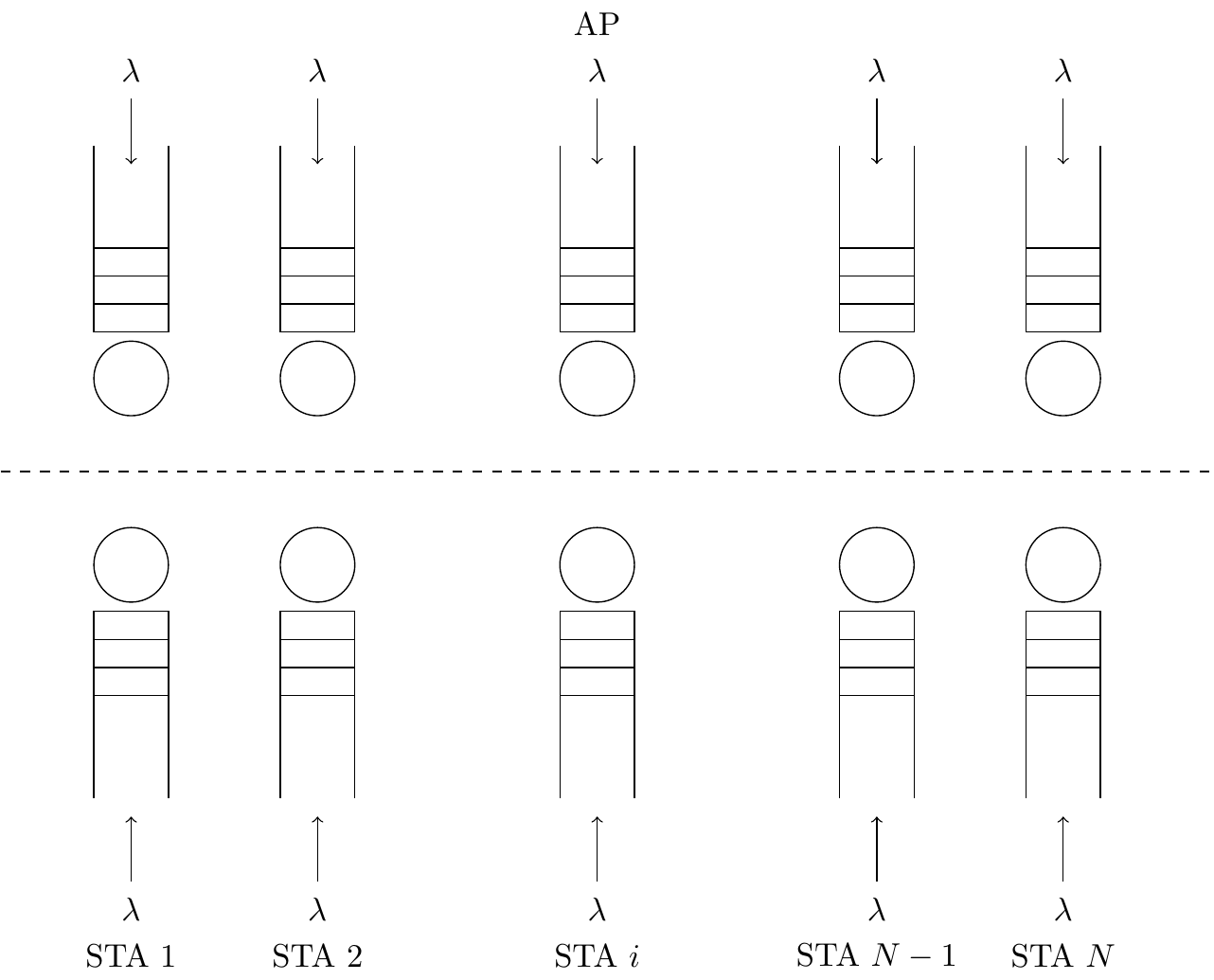}
\caption{System Model for Opportunistic Scheduling}
\label{fig:sys_opp}
\end{figure}

 We consider an infrastructure 802.11 WLAN with $N$ mobile stations (STAs) associated with a single access point (AP) (Fig.~\ref{fig:sys_opp}).
 Each STA has an uplink queue for the packets destined to the AP. 
In the 802.11 standard, the AP has one queue to transmit all data to the users. 
Also, the backoff timers at the AP as well as the STAs use the same algorithm resulting in equal opportunity to access the channel for the AP as well as the STAs.
Since AP has downlink data for all the users, this causes severe bottleneck at the AP queue. 
To handle this, in our setup, the AP has a separate downlink queue for each of the mobile stations.
Packets at the AP are sorted based on their destinations into these queues and each AP queue sets a separate backoff timer.
Hence, the AP in effect sets a backoff timer equivalent to the minimum of all the AP queue backoff timers. 
If more than one AP queues set the minimum value of the timers, the AP randomly selects one among them for transmission in the event that it wins contention. 
Hence there are no collisions among the individual queues at the AP.

We assume that the packets are arriving at each of the queues in the uplink and the downlink as a Poisson process with rate $\lambda$ packets/sec. All packets are of same length.

The channel gains are assumed to be constant over the duration of the transmission of a packet.
 The channel gain from the AP to each of the STAs is assumed to be independent of each other and independent, identically distributed ($\iid$) in time. 
 The independence of the channels is justified as the STAs are assumed to be placed sufficiently far apart from each other. 
The $\iid$ assumption in time is used to make the analysis tractable. However, this is also reasonable and is usually made in literature.

Our scheme is also applicable for networks with non $\iid$ channels, but these assumptions are made for theoretical analysis to be provided later on.

 IEEE 802.11 standards specify multiple modulation and coding schemes at the PHY layer which can be used by a node depending on the instantaneous link quality to the destination to satisfy the desired error performance of the system.  
We can quantize the continously varying channel into a discrete number of states based on the transmission rates used during transmission.

We assume that each queue has perfect channel knowledge for its channel.
Whenever the channel is free for a specified duration (DIFS), the queues with data to transmit contend to access the channel.
In the 802.11 standard, the backoff timer of a user is independent of its channel state.
However, in our algorithm, each of these queues sets a backoff timer which is a function of the instantaneous channel quality to its destination. 
The queue which sees the best channel to its destination will set the smallest timer and will expire first and win the contention. The winning queue now transmits its packet to the destination.
The scheduling algorithm and the timer scheme is explained in detail in the next section.

\section{Timer function}
\label{sec:timerfunc}
 To begin with, the backoff timer set by a queue during the contention period must be a monotonically nonincreasing function of the channel gain to its destination. This ensures that during each contention period, the queue with the best channel to the destination is selected for transmission.
 
 Also, the granularity of the backoff timers set is determined by a parameter called the vulnerability window $\Delta$ of the WLAN that depends on the maximum transmission duration and maximum propagation distance of the network\cite{virag10}.
 If two timers expire within this duration, then their transmissions may result in a collision.

 If both an STA queue and the corresponding queue at the AP  have data to transmit, then due to the electromagnetic reciprocity of the channel, both the STA and the corresponding AP queue for that particular STA will observe the same channel gain to their destinations.
 Therefore, they will set the same value of backoff timer in each contention period.
 This will result in repeated collisions whenever their backoff timers expire before others until the packet is dropped after the retry limit for the packet is reached.
Therefore, we need to include randomness also in the backoff timer.
%

 In particular, if the channel is in state $i$, then the AP and STA queues will set a backoff timer $T_i$  given by,

a) \emph{AP queue}:

\begin{equation}
T_i=
\begin{cases}
2(|\mathcal{H}|-i)\Delta  \text{  $\mu s$}, &\text{$\wpr$ $p$},\\
(2(|\mathcal{H}|-i)+1)\Delta  \text{  $\mu s$}, &\text{$\wpr$ $1-p$}.
\end{cases}
\label{eqn:APtimer}
\end{equation}

b)\emph{STA queue}
\begin{equation}
T_i=
\begin{cases}
2(|\mathcal{H}|-i)\Delta  \text{  $\mu s$}, &\text{$\wpr$ $1-p$},\\
(2(|\mathcal{H}|-i)+1)\Delta  \text{  $\mu s$}, &\text{$\wpr$ $p$}.
\end{cases}
\label{eqn:STAtimer}
\end{equation}
where $|\mathcal{H}|$ is the number of channel states.

As a result, the probability of collision between the timers at an AP queue and the corresponding STA queue is reduced.
The probability of the collision depends on the probability $p$ in equations (\ref{eqn:APtimer}) and (\ref{eqn:STAtimer}).

In the rest of the paper, we analyse the performance of this algorithm and then compare with simulations and the DCF algorithm of 802.11.

\section{Performance Analysis}
For simplicity we assume symmetric statistics for all users: same Poisson arrival rates and same channel gain statistics.
Even then, the exact analysis of the system is intractable and hence an approximate analysis will be attempted.
However, we will show that it approximates well the actual system simulated via Qualnet.

Let $\pap$ and $\psta$ denote the long term probability that an AP queue and a STA queue is nonempty respectively. Because of symmetry, these probabilities will be same for every AP and STA queue respectively.

Let the instants $T_k$, $k \in0,1,2, 3, . . . $ represent the instants when the $k$th successful transmission ends. Let $R_k=T_k-T_{k-1}, k \in 1, 2, 3, ...$ .
We assume that $\{R_k, k \geq 1\}$ are $\iid$.
In other words, the number of packets successfully transmitted is modelled as a Renewal Process, with
the end of successful transfers being the renewal instants.

\subsection{Rate Equations}
\label{sec:fixedpoint}

Consider a particular AP queue and a STA queue pair for a particular node index. In any renewal cycle, though there may be multiple idle or collision events, by definition, there is \emph{exactly} one successful packet transmission. 
Let $\peqnap$ and $\peqnsta$ represent the probability that this successful transmission was by the tagged AP queue and STA queue respectively.

For every renewal cycle, we associate a reward, $I_{AP}=1$ if the tagged AP queue transmitted successfully in this cycle and  $0$ otherwise. Let $H(t)$ represent the number of packets transmitted by the tagged AP queue
 in the interval $[0,t)$. Hence, by renewal reward theorem (\cite{wolf89}), we get the throughput $\Theta_{AP}$ of the AP queue as,
\begin{equation}
 \Theta_{AP}=\lim_{t\to\infty} \frac{H(t)}{t} = \frac{\mathbb{E}[I_{AP}]}{\mathbb{E}[\renew]} = \frac{\peqnap}{\mathbb{E}(\renew)}\text{ a.s. }
\end{equation}
where the random variable $\renew$ represents the length of a renewal cycle.

Similarly, we obtain, the throughput $\Theta_{STA}$ of the STA queue as,
 \begin{equation}
 \Theta_{STA} = \frac{\peqnsta}{\mathbb{E}[\renew]} \text{ a.s. }
\end{equation}
 
  For low values of arrival rate, the queues are stable, i.e the departure rate is equal to the arrival rate, i.e.,

\begin{equation}
 \lambda = \Theta_{AP}\text{      and     }\lambda = \Theta_{STA}. 
 \label{eqn:fixedpoint}
\end{equation}

Assuming that the queues are stable, we solve the set of equations (\ref{eqn:fixedpoint}) for $\pap$ and $\psta$ for increasing values of arrival rate $\lambda$ until the probabilities $\pap$ and $\psta$  are close to $1$. 
This gives us a measure of the capacity of the system. 
In Sections \ref{sec:renewal} and \ref{sec:fixednum}, we derive expressions for $\mathbb{E}(\renew)$ and $\peqnsta$ respectively.
Finally, in Section \ref{sec:oppsimln}, we compare the values of $\pap$ and $\psta$ obtained from analysis above to those from simulation.

\section{Expected Renewal Length}
\label{sec:renewal}
\begin{figure}[h]
 \centering
  \includegraphics[scale=0.6]{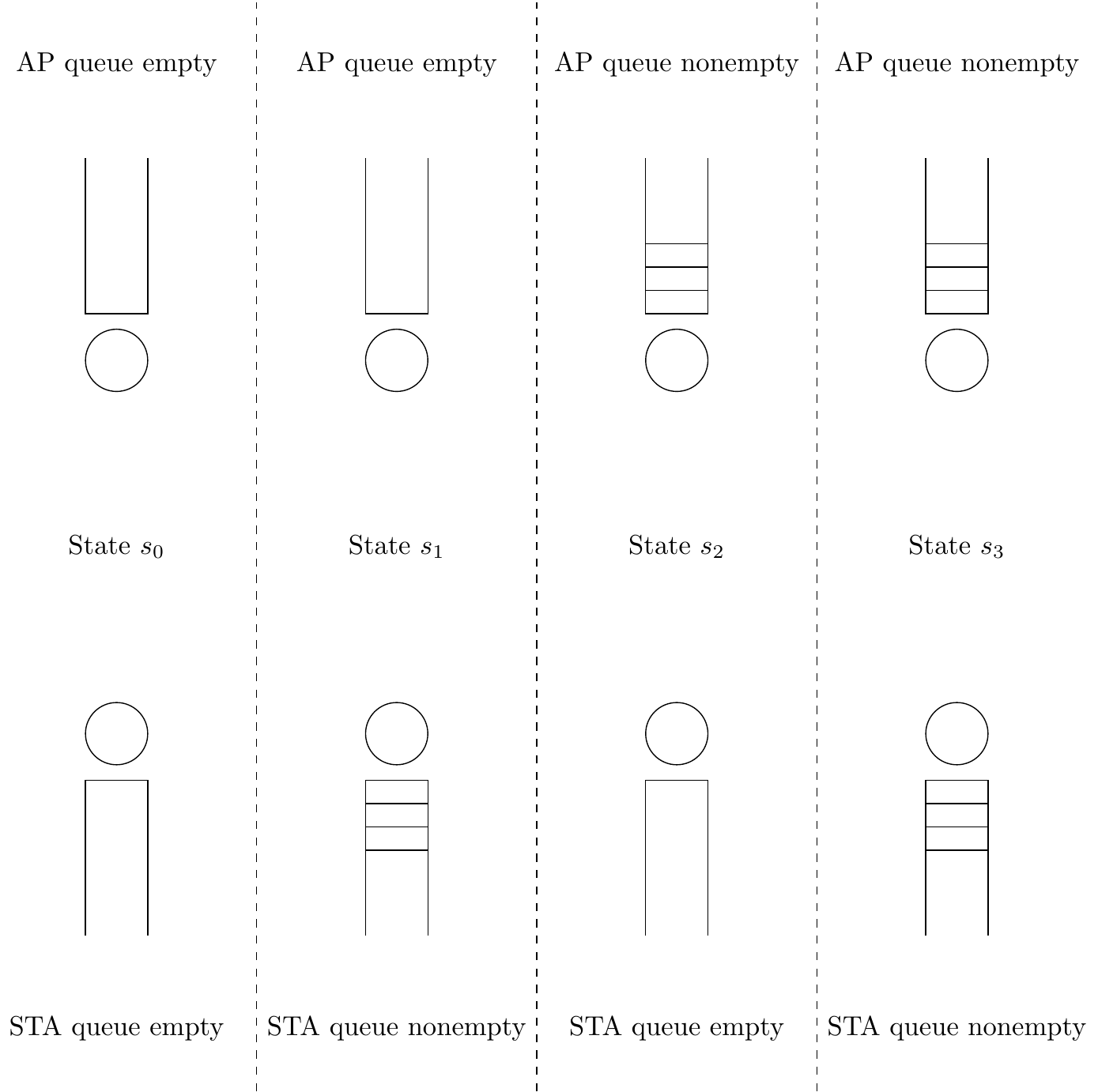}
 \caption{Queue Pair States}
 \label{fig:qpairstates}
 \end{figure}

 First, let us focus on a single AP and STA queue pair. At any time instant, depending on whether the AP and/or the STA queue is empty/nonempty we define the queue pair to be in the either of the four states $\{s_0,...,s_3\}$ 
(Fig \ref{fig:qpairstates}).
       
  The length of a renewal cycle depends on the state of all the queue pairs in the system at the beginning of the cycle. 
  Hence, to calculate expected renewal cycle length $\mathbb{E}[\renew]$, we will condition on the number of queue pairs in each of the possible queue pair states.
At the beginning of each contention period, we keep track of the state $\state=(k_1, k_2, k_3)$  where,

$k_1$ = Number of queue pairs in state $s_1$,

$k_2$ = Number of queue pairs in state $s_2$,

$k_3$ = Number of queue pairs in state $s_3$.

Let $\renew^{(k_{1},k_{2},k_{3})}$ represent the length of the renewal cycle given that the system is in state $\state=(k_{1},k_{2},k_{3})$ at the start of the cycle. Then,

  \begin{equation}
    \EX[\renew] = \sum\limits_{k_{1}=0}^N\sum\limits_{k_{2}=0}^{N-k_{1}}\sum\limits_{k_{3}=0}^{N-k_{1}-k_{2}}\mathbbm{P}\{\state=(k_{1},k_{2},k_{3})\}\EX[\renew^{(k_{1},k_{2},k_{3})}],
\label{eqn:exprenewal2}
   \end{equation} 
  where,
\begin{equation}
 \begin{IEEEeqnarraybox}{rl}  
&\textstyle\mathbbm{P}\{\state=(k_{1},k_{2},k_{3})\}=\binom{N}{k_1}\binom{N-k_1}{k_2}\binom{N-k_1-k_2}{k_3}\\
                       &.P_{AP}^{{k_1+k_3}}(1-P_{AP})^{{N-k_1-k_3}}.P_{STA}^{{k_2+k_3}}(1-P_{STA})^{{N-k_2-k_3}}.
 \end{IEEEeqnarraybox}
\end{equation}

In any renewal cycle, there could be idle slots, one or more collision events followed by a successful transmission by either an AP or a STA queue.
 We can think of the renewal cycle as composed of one or more minislots, where each minislot is defined as the time interval between end of an idle, collision or  packet transmission duration.
 In the following, we write recursive equations in terms of conditional expectations $\renew^{(k_{1},k_{2},k_{3})}$ for all possible system states $(k_{1},k_{2},k_{3})$ based on the state of the system at 
the beginning and the end of the first minislot. 
We solve this set of linear equations and use equation \ref{eqn:exprenewal2} to obtain the expected renewal cycle length.

 \emph{1) CASE 1: $(k_1,k_2,k_3)=(0,0,0)$:}

  If all the queues are empty, i.e., the system is in state $\state=(0,0,0)$, then the system is idle until a packet arrives at any of the queues.
 Let the r.v $\idletime$ represent this duration. It is exponentially distributed with parameter $2N\lambda$.

The first packet which arrives to an empty system is equally likely to arrive at an AP queue or an STA queue. If the packet arrives to an AP queue, the system goes into the state $\state=(1,0,0)$, otherwise it goes to the 
state $\state=(0,1,0)$.
 Therefore,
   \begin{equation}
    \renew^{(0,0,0)}= \idletime + 0.5\renew^{(1,0,0)} + 0.5\renew^{(0,1,0)},
   \end{equation}
and
\begin{equation}
    \mathbb{E}[\renew^{(0,0,0)}]= \frac{1}{2N\lambda} + 0.5\mathbb{E}[\renew^{(1,0,0)}] + 0.5\mathbb{E}[\renew^{(0,1,0)}].
\label{eqn:renewalempty}
   \end{equation}

\subsubsection{CASE II :$(k_1,k_2,k_3)\neq(0,0,0)$}
As explained in section \ref{sec:timerfunc}, $\Delta$ is the vulnerability window of the network and $\delta$ is the 802.11 system slot length. In the following, we assume $\Delta=\delta$, in our setup.
However, the analysis can easily be generalized to an arbitary value of the vulnerability window $\Delta$.

 A contention period begins when the channel is free for more than a stipulated amount of time called the DIFS whose value is defined in the 802.11 standard. Let the time instant $\tau$ represent the beginning of the contention period. 
 If a queue is nonempty at time $\tau$, then it immediately sets its backoff timer as a funtion of the channel gain to its destination. Otherwise, if a packet arrives to an empty queue, say in the  $m^{th}$ slot from the beginning of the contention period,
 then the queue will set its backoff timer at time $\tau+m\delta$. 

  Let $\psucap\big(k;l;(k_{1},k_{2},k_{3})\big)$ and similarly $\psucsta\big(k;l;(k_{1},k_{2},k_{3})\big)$ denote the probability that an AP queue and an STA queue respectively of a queue pair in state $s_i, i=0,.., 3$ wins contention after $k$ slots, $k=0,1, .., T_{max}$ ,
 i.e at time $\tau+k\Delta$ by setting a backoff timer of length $l\Delta$, $l=0, .. , k$ .. Once a queue wins contention, the time taken to transmit a packet depends on the rate of transmission of the packet  which in turn depends on the channel gain it sees to its destination 
which is inversely related to the duration of the backoff timer set.
 
 
 Also let $\pcol\big(k;(k_{1},k_{2},k_{3})\big)$ denote the probability of a collision between two or more queues after $k$ slots, $k=0,1, .., T_{max}$  given that the system in state $\state=(k_{1},k_{2},k_{3})$.
Expressions for the above defined probabilities are derived in appendix \ref{app_oppsched1}-\ref{app_oppsched3}.

    If a successful contention resolution takes place during the contention period, then the winning queue transmits its packet to its destination. If the packet is received correctly, then the renewal cycle ends. Otherwise, if the packet is received in 
error or if a collision occurs during contention, then a new contention period starts when the channel is 
sensed to be free for more than the DIFS period. 
  
 Therefore, for each valid state $\state=(k_1, k_2, k_3)$, we write recursive expressions for $\EX[\renew^{(k_{1},k_{2},k_{3})}]$ considering all the state transitions that can occur following a collision or a failed transmission in equation \ref{eqn:renewrecurse} in page \pageref{eqn:renewrecurse}.
 
 \begin{figure*}

\begin{equation}
\begin{IEEEeqnarraybox}{rll}
\EX[\renew^{(k_{1},k_{2},k_{3})}]=&\sum\limits_{k=0}^{T_{max}}\sum\limits_{l=0}^{k}\sum\limits_{i=0}^{3}(\psucap+\psucsta)\big(k;l;(k_{1},k_{2},k_{3})\big)\cdot(1-\perr(\scriptstyle{\lfloor\frac{T_{max}+1-l}{2}\rfloor}\textstyle))\cdot\\
&\underbrace{\bigg[k\Delta + \tsuc(\scriptstyle{\lfloor\frac{T_{max}+1-l}{2}\rfloor}\textstyle)\bigg]}_\text{Successful contention resolution and transmission} +\\
\\
   &\sum\limits_{k=0}^{T_{max}}\sum\limits_{l=0}^{k}\sum\limits_{i=0}^{3}(\psucap+\psucsta)\big(k;l;(k_{1},k_{2},k_{3}))\cdot\perr(\scriptstyle{\lfloor\frac{T_{max}+1-l}{2}\rfloor}\textstyle))\cdot\\
  &\bigg[k\Delta +\tsuc(\scriptstyle{\lfloor\frac{T_{max}+1-l}{2}\rfloor}\textstyle)+ \sum\limits_{a=0}^{k_1}\sum\limits_{b=0}^{k_2}\sum\limits_{c=0}^{\scriptstyle{N-k1-k2}}\sum\limits_{d=0}^{\scriptstyle{N-k1-k2-c}}\sum\limits_{e=0}^{\scriptstyle{N-k1-k2-c-d}}\\
    &\underbrace{P\{\scriptstyle{(k_1;k_2;k_3)\rightarrow(k_1-a+c;k_2-b+d;k_3+a+b+e)}|t=k\Delta+T_{suc}({\lfloor\frac{T_{max}+1-l}{2}\rfloor}\textstyle)\textstyle\}{\EX[\renew^{(k_{1}-a+c,k_{2}-b+d,k_{3}+a+b+e)}}]\bigg]}_\text{Successful contention resolution and failed transmission}\\ 
\\
     &+\sum\limits_{k=0}^{T_{max}}\pcol\big(k;(k_{1},k_{2},k_{3}))\cdot\bigg[k\Delta +\tcol+ \sum\limits_{a=0}^{k_1}\sum\limits_{b=0}^{k_2}\sum\limits_{c=0}^{\scriptstyle{N-k1-k2}}\sum\limits_{d=0}^{\scriptstyle{N-k1-k2-c}}\sum\limits_{e=0}^{\scriptstyle{N-k1-k2-c-d}}\\
    &\underbrace{P\{\scriptstyle{(k_1;k_2;k_3)\rightarrow(k_1-a+c;k_2-b+d;k_3+a+b+e)|t=k\Delta+T_{col}}\textstyle\}{\EX[\renew^{(k_{1}-a+c,k_{2}-b+d,k_{3}+a+b+e)}}]\bigg]}_{Collision} 
\end{IEEEeqnarraybox}
\label{eqn:renewrecurse}
     \end{equation}

where,

 
%
 \begin{align*}
T_{col}&-\text{Time spent in collision.}\\
T_{suc}(i)&-\text{Time spent in transmission of a packet given the channel is in state $i$.}\\
\perr(i)&-\text{Probability of transmission error given the channel is in state $i$.}\\
P\{\scriptstyle{(k_1;k_2;k_3)\rightarrow(m_1;m_2;m_3)}\textstyle|t=T\}& -\text{Probability that the system transitions from state }(k_1,k_2,k_3)\text{ to } (m_1;m_2;m_3)\\ 
 &\hspace{1em}\text{ in a duration of time } T \text{ microseconds.} \\
a&-\text{No. of queue pairs which transition from state $s_1$ to $s_3$.}\\
b&-\text{No. of queue pairs which transition from state $s_2$ to $s_3$.}\\
c&-\text{No. of queue pairs which transition from state $s_0$ to $s_1$.}\\
d&-\text{No. of queue pairs which transition from state $s_0$ to $s_2$.}\\
e&-\text{No. of queue pairs which transition from state $s_0$ to $s_3$.}
\end{align*}

Therefore,
\begin{equation}
 \begin{IEEEeqnarraybox}{rCl}
 P\{\scriptstyle{(k_1;k_2;k_3)\rightarrow(k_1-a+c;k_2-b+d;k_3+a+b+e)}\textstyle|t=T\}&=&\binom{k_1}{a}\binom{k_2}{b}\binom{\scriptstyle N-k_1-k_2}{c}\cdot\\
&&\binom{\scriptstyle N-k_1-k_2-k_3-c}{d}\binom{\scriptstyle N-k_1-k_2-k_3-c-d}{e}\cdot\\
&& P_{Nonempty}(T)^{a+b+c+d+2e}\cdot\\
&&(1- P_{Nonempty}(T))^{\scriptstyle k_1-a+k_2-b+c+d+2(N-k_1-k_2-k_3-c-d-e)}
  \end{IEEEeqnarraybox}
\end{equation}

\begin{equation}
\begin{IEEEeqnarraybox}{rCl}
 P_{Nonempty}(T)&=& P\{\text{\footnotesize Atleast one packet arrives in the duration $(t,t+T]|$Queue is empty at time $t$}\}\\
                &=& 1-e^{-\lambda T} 
  \end{IEEEeqnarraybox}
\end{equation}
     \line(1,0){520}
     \end{figure*}
     
\section{Calculation of $\peqnap$ and $\peqnsta$}
\label{sec:fixednum}

	To calculate $\peqnap$ and $\peqnsta$, we follow the same recursive approach as in Section \ref{sec:renewal}.
 However, unlike in Section \ref{sec:renewal}, we now keep track of the status of the tagged queue pair (section \ref{sec:fixedpoint}) and the remaining $N-1$ queue pairs separately at the beginning of each contention period.
In this section, we refer to the state of the system at any given time as represented by the  four-tuple $(s_i; l_1,l_2, l_3)$  
 where,
	
	$s_i$ - state of the tagged queue pair,
	
	$l_1$ - Number of nontagged queue pairs in state $s_1$,
	
	$l_2$ - Number of nontagged queue pairs in state $s_2$,
	
	$l_3$ - Number of nontagged queue pairs in state $s_3$.
	
Let $\pcondeqnap$ and $\pcondeqnsta$ denote the probability that the AP queue and the STA queue of the tagged queue pair wins contention and transmits successfully in a renewal cycle given that the contention period began in the state $(s_i, l_1, l_2, l_3)$.

Then,
	\begin{equation}
\begin{IEEEeqnarraybox}{rlll} 
 \peqnap&=&\sum\limits_{i=0}^{3}\sum\limits_{l_1=0}^{N-1}\sum\limits_{l_2=0}^{N-1-l_1}\sum\limits_{l_3=0}^{N-1-l_1-l_2}&\mathbbm{P}\{\state=(s_i;l_{1},l_{2},l_{3})\}\\
         &&&\cdot\pcondeqnap
\label{eq:pcondap}
\end{IEEEeqnarraybox}
\end{equation}
	
	and
	
	\begin{equation}
\begin{IEEEeqnarraybox}{rlll} 
 \peqnsta &=&\sum\limits_{i=0}^{3}\sum\limits_{l_1=0}^{N-1}\sum\limits_{l_2=0}^{N-1-l_1}\sum\limits_{l_3=0}^{N-1-l_1-l_2}&\mathbbm{P}\{\state=(s_i;l_{1},l_{2},l_{3})\}\\
         &&&\cdot\pcondeqnsta,
\end{IEEEeqnarraybox}
\label{eq:pcondsta}
\end{equation}
where $\mathbbm{P}\{\state=(s_i;l_{1},l_{2},l_{3})\}$ is the probability that the system is in state $\state=(s_i;l_{1},l_{2},l_{3})$.

 As in section \ref{sec:renewal} we divide into multiple minislots where each minislot is the time interval between end of idle interval, collisions or transmissions.
 
 Let $\papwinms$ and $\pstawinms$ denote the probability that the tagged AP queue and tagged STA queue win contention and transmit successfully in a minislot given that the system is in state $(s_i, l_1,l_2, l_3)$ at the beginning of the minislot.
Expressions for the above defined probabilities are derived in appendix \ref{app_oppsched4}.

 By conditioning on the state of the system at the beginning of the minislot we write in equation \ref{eqn:approbrecurse} in page \pageref{eqn:approbrecurse}.
 In the same manner, $\pstawinms$ is also obtained.
\begin{figure*}
\begin{equation}
	\begin{IEEEeqnarraybox}{rll} \pcondeqnap=&\papwinms+\sum\limits_{k=0}^{T_{MAX}}\sum\limits_{j=0}^{3}\sum\limits_{a=0}^{l_1}\sum\limits_{b=0}^{l_2}\sum\limits_{c=0}^{\scriptstyle{N-l_1-l_2}}\sum\limits_{d=0}^{\scriptstyle{N-l_1-l_2-c}}\sum\limits_{e=0}^{\scriptstyle{N-k1-k2-c-d}}\pcol(k;s_i;l_1,l_2,l_3)\cdot\\
&\underbrace{P\{\scriptstyle{(s_i;l_1,l_2,l_3)\rightarrow(s_j;l_1-a+c,l_2-b+d,l_3+a+b+e)|t=T_{col}+k\Delta}\textstyle\} \pcondeqnapend}_\text{Collision}+\\
&\sum\limits_{k=0}^{T_{MAX}}\sum\limits_{m=0}^{k}\sum\limits_{n=0}^{3}\sum\limits_{j=0}^{3}\sum\limits_{a=0}^{l_1}\sum\limits_{b=0}^{l_2}\sum\limits_{c=0}^{\scriptstyle{N-l_1-l_2}}\sum\limits_{d=0}^{\scriptstyle{N-l_1-l_2-c}}\sum\limits_{e=0}^{\scriptstyle{N-k1-k2-c-d}}\\
&(P_{suc, AP}^{n}+P_{suc, AP}^{n})\big(k;m;(s_i;l_{1},l_{2},l_{3}))\cdot\\
&\underbrace{P\{\scriptstyle{(s_i;l_1,l_2,l_3)\rightarrow(s_j;l_1-a+c,l_2-b+d,l_3+a+b+e)|t=T_{suc}({\lfloor\frac{T_{max}+1-m}{2}\rfloor})+k\Delta}\textstyle\} \pcondeqnapend}_\text{Failed transmission}.\\
 \end{IEEEeqnarraybox}
\label{eqn:approbrecurse}
\end{equation}

\line(1,0){520}
\end{figure*}
 
\section{Simulation and numerical results}
\label{sec:oppsimln}
In this section, we compare the values of  $P_{AP}$ and $P_{STA}$ from simulation to those obtained from analysis.
 We further compare the performance of our opportunistic scheduling scheme against the standard 802.11a protocol.

In \cite{wang2009joint}, Wang obtain $\frac{E_b}{N_0}$ (the energy per bit to noise power spectral density ratio) thresholds to maintain a packet error rate (PER) of $0.1$ in transmitting a $1500 Byte$ payload for each data rate supported by the 802.11a MAC. 
Though the 802.11a standard provides 8 PHY modes, the authors in \cite{wang2009joint} argue that it is sufficient to consider the modes with data rates $12$, $24$, $48$ and $54$ Mbps for the purpose of rate adaptation.  
Accordingly, we remove the inefficient PHY modes and quantize the channel into $4$ states corresponding to rates $12$, $24$, $48$ and $54$ Mbps respectively.
The quantization region for each of the data rates is given in Table \ref{tab:quantization}.

\begin{table}[!ht]
\centering
\caption{$\frac{E_b}{N_0}$ range for each channel state with corresponding data rate to maintain PER of $0.1$ for a 1500B payload. }
\begin{tabular}{|c|c | c | }
\hline
Channel State &$\frac{E_b}{N_0}$ Range &Data Rate\\
\hline
$H_0$ &$[0,19.11)$ &12 Mbps\\
$H_1$ &$[19.11,26.90)$ &24 Mbps\\
$H_2$ &$[26.90,31.88)$ &48Mbps\\
$H_3$ &$[31.88, \infty)$ &54Mbps\\
\hline
\end{tabular}
\label{tab:quantization}
\end{table}

Simulations were performed using the Qualnet v4.5 network simulator which has an accurate 802.11a MAC implementation.
The Qualnet libraries were modified to implement the opportunistic scheme. We took $p$ in equations (\ref{eqn:APtimer}) and (\ref{eqn:STAtimer}) to be $0.5$. The simulation setup consists of $7$ STAs associated with an AP. 
The STAs were placed at an equal distance of $100m$ from the AP to remove differences in large scale fading. We  use the free space path loss model with no shadowing. 

\begin{figure*}
\null\hfill
\hspace{-1em}
\subfloat[][\label{theorysimln:a} $\pap$ and $\psta$]{
 \includegraphics[width=.5\textwidth]{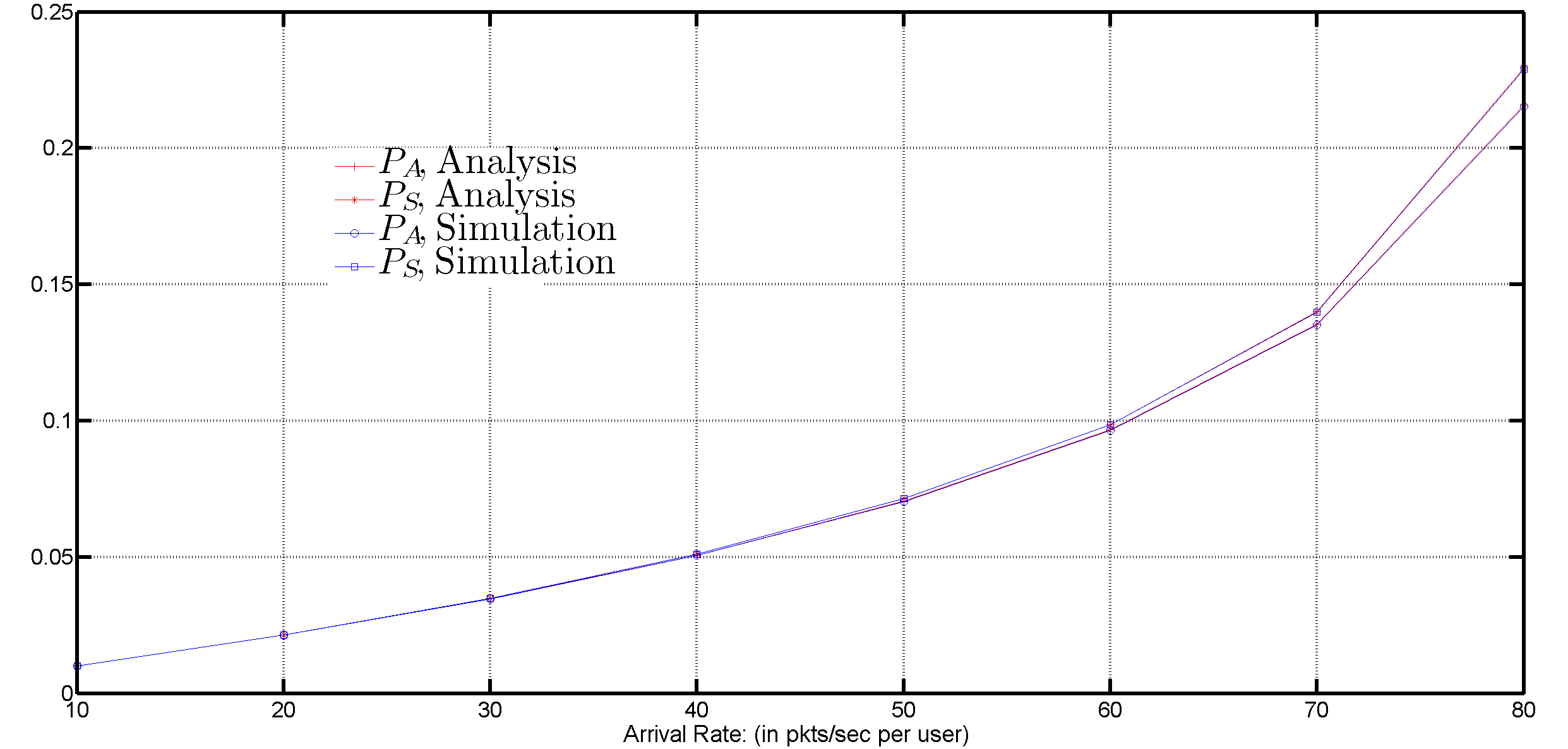}
 }
\subfloat[][\label{theorysimln:b} Expected Renewal Cycle Length, $\mathbb{E}{[\renew]}$]{
 \includegraphics[width=.5\textwidth]{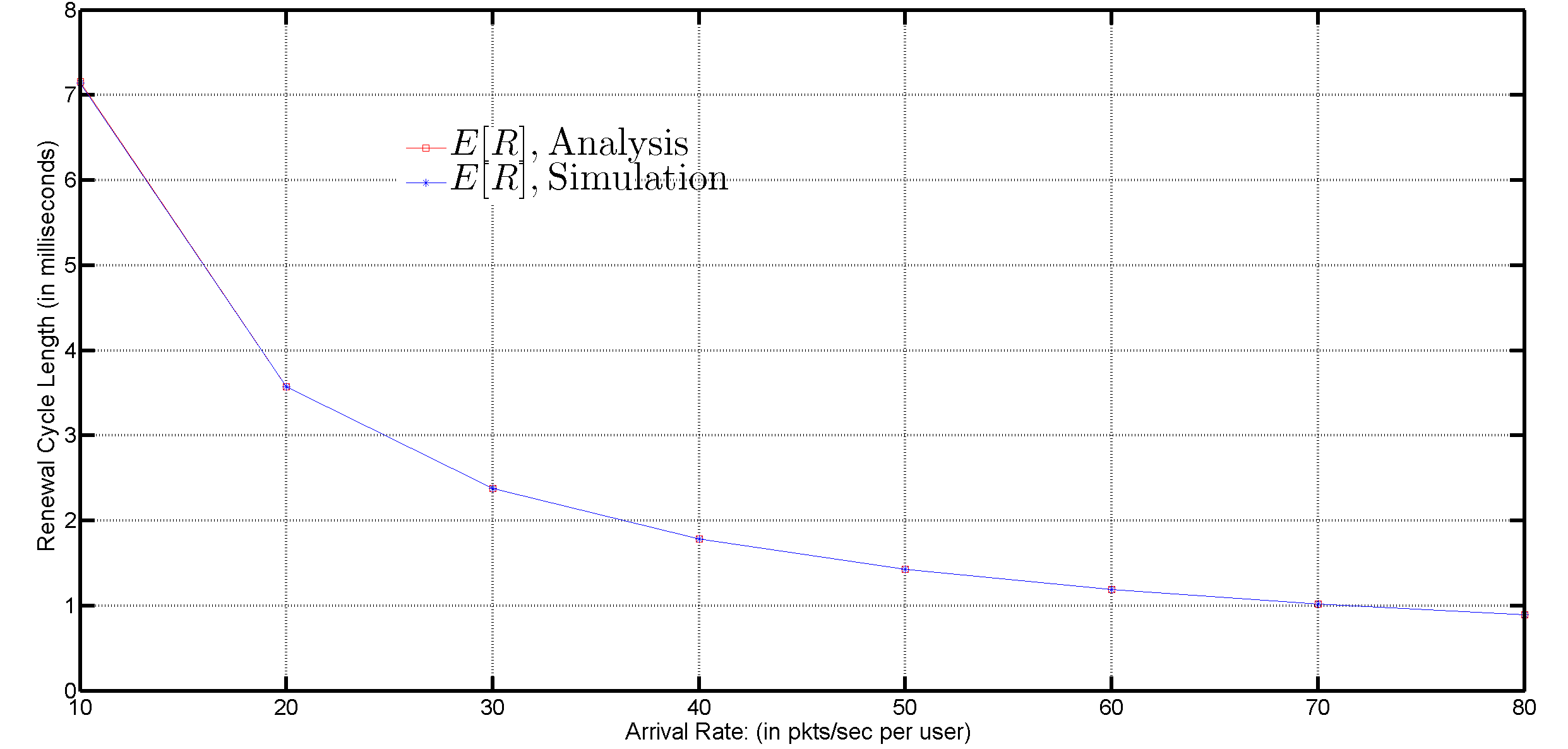}
}
\hfill\null
\caption{ comparision of simulations and theory.}
\label{fig:theorysimln}
\end{figure*}

\begin{figure*}[!t]
\null\hfill
\hspace{-1em}
\subfloat[][\label{oppsimln:a} Aggregate downlink throughput]{
\includegraphics[width=.5\textwidth]{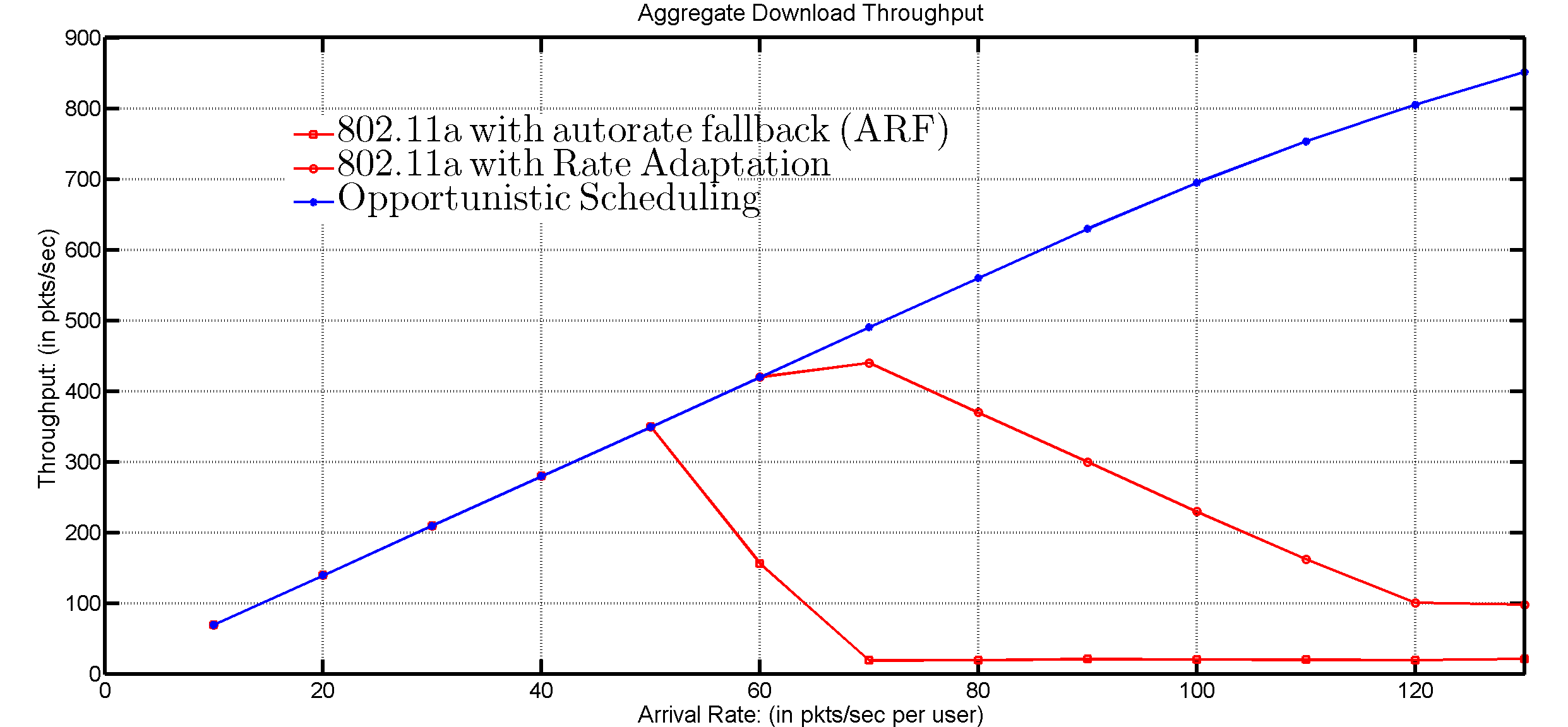}
}
\subfloat[][\label{oppsimln:b} System throughput]{
\includegraphics[width=.5\textwidth]{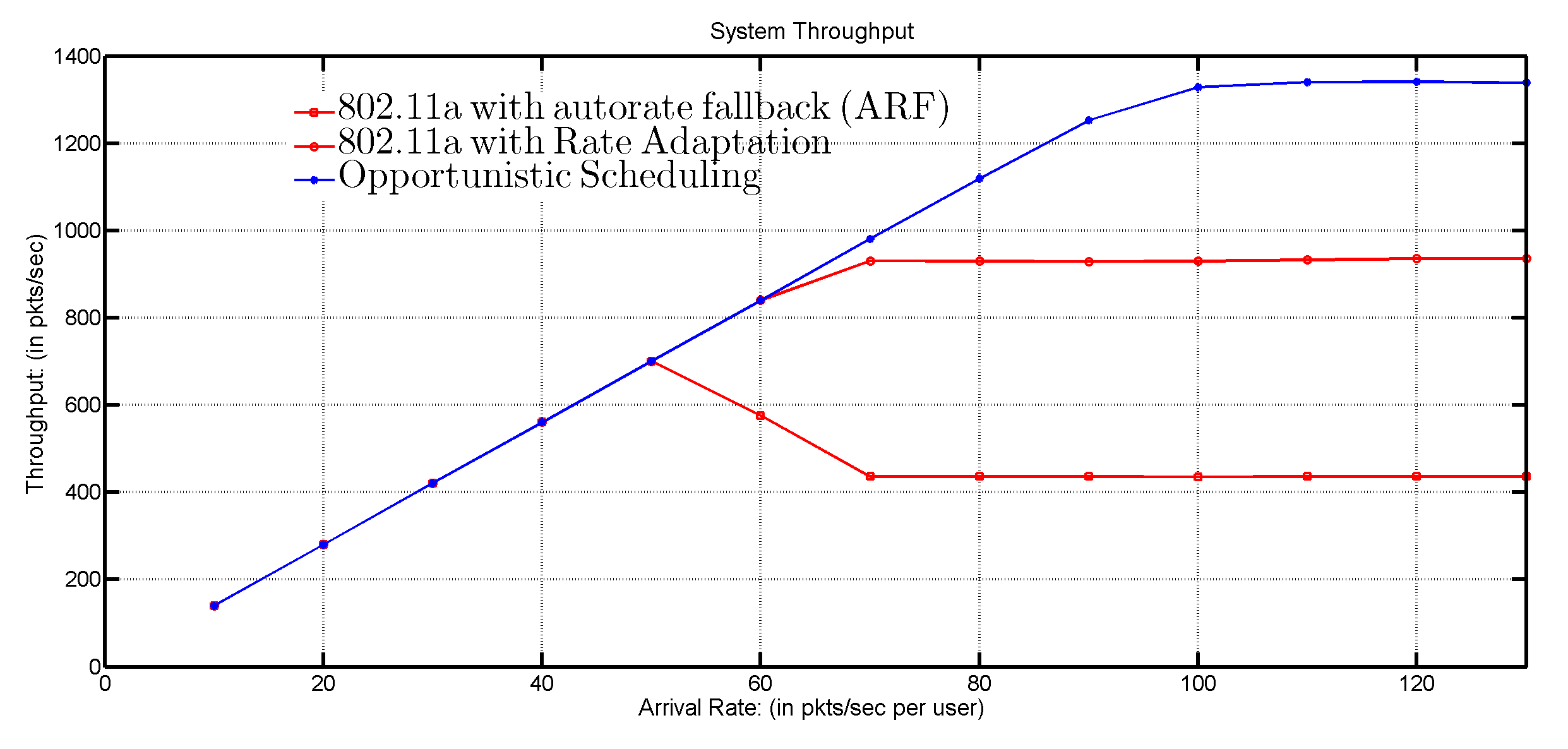}
}
\hfill\null
\caption{Throughput performance of opportunistic scheduling}
\label{fig:oppsimln}
\end{figure*}

To begin with, we consider a simplified error model and set $PER=0.1$ for all users. We compare the probabilities $\pap$ and $\psta$ and expected cycle length $\mathbb{E}[\renew]$ obtained by analysis to those obtained by simulation (Fig \ref{fig:theorysimln}).
We see the analysis to be remarkably accurate in predicting the behavior of the system. During simulations we observed that as the arrival rate $\lambda\to90$, $\psta\to 1$ which agrees with the analytical results.
 It was  observed that when $\lambda\geq90$  pkts/sec, the fixed point iteration to solve equation (\ref{eqn:fixedpoint}) fails to converge as the queues become unstable. 

Now, we compare the throughput of our algorithm with the 802.11a DCF protocol in a Rayleigh fading environment. 
Hence, the SNR of the received signal is exponentially distributed with mean $\overline{SNR}$ which is same for all nodes as the STAs are equidistant from the AP.

The BER for a given SNR is obtained using precomputed values from 
BER lookup table implemented in Qualnet. 

  Though the 802.11 standard specifies multiple modulation and coding schemes for rate adaptation 
at the PHY layer, the choice of the appropriate rate for a given channel conditions is left to the user. 
One of the earliest and widely deployed rate adaptation algorithms in the 802.11 standard is the Autorate fallback (ARF) scheme. In ARF, the sender starts transmitting at the lowest data rate.
 After a fixed number of successful transmissions (say $10$) at a given rate, the sender attempts to use a higher transmission rate. Similarly, the sender falls back to a lower transmission rate after two successive failures.

The choice of the rate adaptation algorithm has a major impact on the performance of the network.
We compare the throughput of the opportunistic scheduling scheme with the rate adaptation algorithm in Table \ref{tab:quantization}  and compare with the throughput of the DCF protocol with ARF as well as with
the adaptation algorithm in \ref{tab:quantization}.
The comparision is provided in Fig~\ref{fig:oppsimln}.

 We see that, both opportunistic scheduling and 802.11a DCF have similar throughput performance for low values of the packet arrival rate (all the queues are stable).
 As the arrival rate increases, the download throughput of the AP in 802.11a (Fig \ref{oppsimln:a}) decreases rapidly at about $50$ pkts/sec (per user) for ARF scheme and at $60$ pkts/sec (per user) for the rate adaptation scheme.
 In contrast, we see a steady increase in download throughput with respect to the arrival rate in the case of opportunistic scheduling.  

 From Fig \ref{oppsimln:b} we observe that the queues are stable upto $50$ pkts/sec (per user) for DCF with ARF scheme and $60$ pkts/sec (per user) for DCF with the rate adaptation scheme mentioned before.
 On the other hand, the queues are stable for arrival rate as high as $90$ pkts/sec (per user) with the opportunistic scheduling scheme.

 Comparing Fig \ref{oppsimln:a} and \ref{oppsimln:b} we can see that DCF protocol is unfair to the AP while the opportunistic scheme provides almost same throughput in the uplink and downlink.

\section{Conclusions}
\label{sec:conclusion}
 We have proposed a joint uplink and downlink opportunistic scheduling scheme for infrastructure WLANs which addresses the issues of uplink/downlink unfairness and the inefficiency of the MAC protocol. 
The uplink downlink unfairness is taken care of by employing multiple queues and backoff timers at the access point (AP).
The inefficiency is handled by using opportunistic scheduling of the channels.
This is done with minimal changes with the DCF protocol of 802.11.
We have also obtained theoretically the performance of our new protocol.
These closely approximate the performance obtained via Qualnet Simulations.
Also, our results further show large improvements in the system throughput with the new opportunistic scheduling scheme over the conventional IEEE 802.11 DCF scheme.

\appendix
\label{app_oppsched1}
Consider a  queue pair consisting of an uplink queue at an STA and the corresponding queue in the downlink for that STA at the AP.
Let $\tau$ represent the beginning of a contention period. 
If either the AP or/and the STA queue is nonempty  at time $\tau$ then the corresponding queue(s) of the queue pair will set a backoff timer as a function of their channel gain.
 
Given that the queue pair is in state $s_i, i=0,...,3$, define $P^{s_i}(\small{k;l;AP})$ and $P^{s_i}(\small{k;l;STA})$ to be the probability that the AP queue and STA queue respectively sets a timer for duration $l\Delta$ expires first at time $\tau+k\Delta$. 
Also, let $P^{s_i}(k;l;AP+STA)$ represent the probability that both the AP and the STA queue set timers for duration $l\Delta$ and expire together at time $\tau+k\Delta$.  
These probabilities can be calculated using the channel state distribution and the timer function defined in section \ref{sec:timerfunc}.

For each contention period, let the random variable $\tau^i_{min}$  represent the minimum of the AP and STA queue timer expiry duration given that the queue pair is in state $s_i$ at the beginning of the contention period.
$\tau^i_{min}=\infty$ if neither AP nor STA queue sets its backoff timer in a particular contention period.
Therefore,
\begin{equation}
\begin{IEEEeqnarraybox}{rCl}
 P(\tau^i_{min}>k)&= &1 - \sum\limits_{m=0}^k\sum\limits_{n=0}^m\bigg[P^{s_i}(m;n;AP)\\
            &&+P^{s_i}(m;n;STA)+P^{s_i}(m;n;AP+STA)\bigg]
 \end{IEEEeqnarraybox}
\end{equation}

Suppose the system is in state $\state$ at the beginning of the contention period.
Let $n_{i},\text{}i=0,..,3$ represent the number of queue pairs in state $s_i$ given that the system is in state $\state$. 
For instance if $\state=(k_1, k_2, k_3)$, then $n_i=k_i, i=1,2,3$ and $n_0=N-k_1-k_2-k_3$.

We now compute the values of $\psucap(k;l;\state)$, $\psucsta(k;l;\state)$ and $\pcol(k;\state)$ used in section \ref{sec:renewal} in terms of the above probabilities as follows:

\subsection{Computation of  $\psucsta(k;l;\state)$}

For a STA queue of a queue pair in state $s_i$ to win contention after $k$ slots system is in state $\state$:
\begin{itemize}
 \item Backoff timer of  the STA queue of any one of the $n_i$ queue pairs in state $s_i$ must expire first after exactly $k$ slots from the beginning of the contention period.
 \item Backoff timers of all the queues in other queue pairs should not expire within $k$ slots.
\end{itemize}

Therefore for $i=0,..,3$ we write,

\begin{equation}
\begin{IEEEeqnarraybox}{rll}
 P_{suc,STA}^{i}(k;l;\state)&=&n_{i}\cdot P^{s_i}(k;l;STA) \cdot P(\tau^i_{min} >k)^{n_i-1}\\
 &&\cdot\bigg(\prod_{j=0,j\neq i}^{3}P(\tau^j_{min} >k)^{n_{j}}\bigg)\\       
\end{IEEEeqnarraybox}
\end{equation}

\subsection{Computation of  $\psucap(k;l;\state)$}
\label{app_oppsched2}
For the AP to win contention after $k$ slots:
\begin{itemize}
 \item  The backoff timer of one or more of the queues at the AP must expire after $k$ slots.
 \item  All other backoff timer should not expire within $k$ slots.
\end{itemize}

Given that the system is in state $\state$, let $a_i$ represent the number of queue pairs whose AP timer expire first after exactly $k$ slots from the beginning of the contention period.
Among the AP queues which expire after $k$ slots, the AP with uniform probability picks any of these queues as the winning queue.
Hence, the probability that an AP queue in state $s_i$ wins contention after $k$ slots by setting a timer of length $l$ slots is given by,

\begin{equation}
 \begin{IEEEeqnarraybox}{rCl}
\psucap(k;l;\state)&=&\sum\limits_{a_0=0}^{n_0}\sum\limits_{a_1=0}^{n_1}\sum\limits_{a_2=0}^{n_2}\sum\limits_{a_3=0}^{n_3}\frac{a_i}{(\sum_{j=0}^3{a_j})}\prod_{j=0}^3{\binom{n_j}{a_j}}\\
&&n_i \cdot {P^{s_i}(k;l;AP)} \cdot\bigg[\sum\limits_{l=0}^{k}{P^{s_i}(k;l;AP)}\bigg]^{a_i-1}\\
&&\bigg(\prod_{j=0, j \neq i}^{3}\bigg[\sum\limits_{l=0}^{k}{P^{s_j}(k;l;AP)}\bigg]^{a_j}\bigg)\\
&&\bigg(\prod_{j=0}^{3}{P(\tau_{s_j}>k)}^{n_j-a_j}\bigg)\\
  \end{IEEEeqnarraybox}
\end{equation}

\subsection{Computation of $\pcol(k;\state)$}
\label{app_oppsched3}
\begin{equation}
  \begin{IEEEeqnarraybox}{rCl}
   \pcol(k;\state)&=&\bigg(\prod_{i=0}^{3}{P(\tau^i_{min} >k-1)}^{n_i}\bigg)\\
		&&-\bigg(\prod_{i=0}^{3}{P(\tau^i_{min} >k\textstyle)}^{n_i}\bigg)\\
                       &&-\sum\limits_{i=0}^{3}\sum\limits_{l=0}^{k}(\psucap+\psucsta)\big(k;l;\state\big)
  \end{IEEEeqnarraybox}
\end{equation}

The first term on the RHS is the probability that the backoff timers of all the queues  expire after atleast $k-1$ slots from the beginning of the contention period.
The second term is the probability that the backoff timers of all the queues expire after atleast $k$ slots from the beginning of the contention period.
The last term is the probability that there is a successful contention resolution after $k$ slots from the beginning of the contention period.

\subsection{Computation of $\papwinms$ and $\pstawinms$}
\label{app_oppsched4}
In this section we calculate $\papwinms$ and $\pstawinms$ given that the system is in state $(s_i;l_{1},l_{2},l_{3})$ which is used in section \ref{sec:fixednum} to calculate $\peqnap$ and $\peqnsta$.
For the tagged AP to transmit successfully in the minislot, the following two events must occur:
\begin{enumerate}
 \item The tagged  AP queue must win contention. This occurs only if:
 \begin{itemize}
  \item The tagged AP queue sets a backoff timer which expires after $k$ slots, \\
     $k=0,1,...,T_{MAX}$. 
 \item Backoff timers of all nontagged queues must expire after $k$ slots from the beginning of the contention period. 
 \end{itemize}
 \item The transmission from the tagged queue must not be received in error.
\end{enumerate}

Therefore,

\begin{equation}
 \begin{IEEEeqnarraybox}{rCl}
  \papwinms&=&\bigg[\sum\limits_{k=0}^{T_{MAX}}\sum\limits_{l=0}^k P^{s_i}(k;l;AP)\\
&&\cdot(1-\perr\big(\lfloor\scriptstyle{\frac{T_{max}+1-l}{2}}\textstyle\rfloor\big)\big)\bigg]\\
       &&\cdot\textstyle P(\scriptstyle\tau^1_{min} >k-1\textstyle)^{l_1}\cdot P(\scriptstyle\tau^2_{min} >k-1\textstyle)^{l_2}\\
&&\cdot P(\scriptstyle\tau^3_{min} >k-1\textstyle)^{l_3}\cdot P(\scriptstyle\tau^0_{min} >k-1\textstyle)^{N-1-l_1-l_2-l_3}.\\
 \end{IEEEeqnarraybox}
\end{equation}

Similarly we can compute $\pstawinms$.

\bibliographystyle{ieeetr}
\bibliography{mythesis}

\begin{thebibliography}{10}

\bibitem{bianchi00}
G.~Bianchi, ``Performance analysis of the ieee 802.11 distributed coordination
  function,'' {\em Selected Areas in Communications, IEEE Journal on}, vol.~18,
  pp.~535 --547, march 2000.

\bibitem{kumar05}
A.~Kumar, E.~Altman, D.~Mior, and M.~Goyal, ``New insights from a fixed-point
  analysis of single cell,'' in {\em IEEE 802.11 WLANs. In: Proc. IEEE Infocom
  2005}, pp.~1550--1561, 2005.

\bibitem{kuriakose07}
G.~Kuriakose, S.~Harsha, A.~Kumar, and V.~Sharma, ``Analytical models for
  capacity estimation of ieee 802.11 wlans using dcf for internet
  applications,'' {\em Wireless Networks}, vol.~15, pp.~259--277, 2009.

\bibitem{Heusse03}
M.~Heusse, F.~Rousseau, G.~Berger-Sabbatel, and A.~Duda, ``Performance anomaly
  of 802.11b,'' in {\em INFOCOM 2003. Twenty-Second Annual Joint Conference of
  the IEEE Computer and Communications. IEEE Societies}, vol.~2, pp.~836--843
  vol.2, 2003.

\bibitem{kim2005downlink}
S.~W. Kim, B.-S. Kim, and Y.~Fang, ``Downlink and uplink resource allocation in
  ieee 802.11 wireless lans,'' {\em Vehicular Technology, IEEE Transactions
  on}, vol.~54, no.~1, pp.~320--327, 2005.

\bibitem{nischal13}
S.~Nischal and V.~Sharma, ``A cooperative arq scheme for infrastructure
  wlans,'' in {\em Wireless Communications and Networking Conference (WCNC),
  2013 IEEE}, pp.~428--433, 2013.

\bibitem{Ji04}
Z.~Ji, Y.~Yang, J.~Zhou, M.~Takai, and R.~Bagrodia, ``Exploiting medium access
  diversity in rate adaptive wireless lans,'' in {\em Proceedings of the 10th
  annual international conference on Mobile computing and networking}, MobiCom
  '04, (New York, NY, USA), pp.~345--359, ACM, 2004.

\bibitem{wangOSMA}
J.~Wang, H.~Zhai, and Y.~Fang, ``Opportunistic packet scheduling and media
  access control for wireless lans and multi-hop ad hoc networks,'' in {\em
  Wireless Communications and Networking Conference, 2004. WCNC. 2004 IEEE},
  vol.~2, pp.~1234--1239 Vol.2, 2004.

\bibitem{DBLP:conf/ifip6-8/HahmLK06}
S.~il~Hahm, J.~Lee, and C.~kwon Kim, ``Distributed opportunistic scheduling in
  ieee 802.11 wlans,'' in {\em PWC} (P.~Cuenca and L.~Orozco-Barbosa, eds.),
  vol.~4217 of {\em Lecture Notes in Computer Science}, pp.~263--274, Springer,
  2006.

\bibitem{Kim06:37}
S.~W. Kim, ``Opportunistic packet scheduling over ieee 802.11 wlan,'' in {\em
  Ubiquitous Intelligence and Computing, Third International Conference, UIC
  2006, Wuhan, China, September 3-6, 2006, Proceedings} (J.~Ma, H.~Jin, L.~T.
  Yang, and J.~J.~P. Tsai, eds.), vol.~4159 of {\em Lecture Notes in Computer
  Science}, pp.~399--408, Springer, 2006.

\bibitem{cioffiuplink}
C.-S. Hwang and J.~Cioffi, ``Opportunistic csma/ca for achieving multi-user
  diversity in wireless lan,'' {\em Wireless Communications, IEEE Transactions
  on}, vol.~8, no.~6, pp.~2972--2982, 2009.

\bibitem{Yoo08}
J.~Yoo, H.~Luo, and C.-k. Kim, ``Joint uplink/downlink opportunistic scheduling
  for wi-fi wlans,'' {\em Comput. Commun.}, vol.~31, pp.~3372--3383, Sept.
  2008.

\bibitem{DBLP:conf/vtc/DianatiT08}
M.~Dianati and R.~Tafazolli, ``Opportunistic scheduling over wireless fading
  channels without explicit feedback,'' in {\em VTC Spring}, pp.~1871--1875,
  IEEE, 2008.

\bibitem{lopez2008}
E.~Lopez-Aguilera, M.~Heusse, Y.~Grunenberger, F.~Rousseau, A.~Duda, and
  J.~Casademont, ``An asymmetric access point for solving the unfairness
  problem in wlans,'' {\em Mobile Computing, IEEE Transactions on}, vol.~7,
  no.~10, pp.~1213--1227, 2008.

\bibitem{hirantha2008dynamic}
B.~Hirantha Sithira~Abeysekera, T.~Matsuda, and T.~Takine, ``Dynamic contention
  window control mechanism to achieve fairness between uplink and downlink
  flows in ieee 802.11 wireless lans,'' {\em Wireless Communications, IEEE
  Transactions on}, vol.~7, no.~9, pp.~3517--3525, 2008.

\bibitem{gopalakrishnan2004}
P.~Gopalakrishnan, D.~Famolari, and T.~Kodama, ``Improving wlan voice capacity
  through dynamic priority access,'' in {\em Global Telecommunications
  Conference, 2004. GLOBECOM'04. IEEE}, vol.~5, pp.~3245--3249, IEEE, 2004.

\bibitem{keceliweighted}
F.~Keceli, I.~Inan, and E.~Ayanoglu, ``Weighted fair uplink/downlink access
  provisioning in ieee 802.11 e wlans,''

\bibitem{pilosof2003understanding}
S.~Pilosof, R.~Ramjee, D.~Raz, Y.~Shavitt, and P.~Sinha, ``Understanding tcp
  fairness over wireless lan,'' in {\em INFOCOM 2003. Twenty-Second Annual
  Joint Conference of the IEEE Computer and Communications. IEEE Societies},
  vol.~2, pp.~863--872, IEEE.

\bibitem{wu2005upstream}
Y.~Wu, Z.~Niu, and J.~Zhu, ``Upstream/downstream unfairness issue of tcp over
  wireless lans with per-flow queueing,'' in {\em Communications, 2005. ICC
  2005. 2005 IEEE International Conference on}, vol.~5, pp.~3543--3547, IEEE,
  2005.

\bibitem{ha2006wlc29}
J.~Ha and C.-H. Choi, ``Wlc29-5: Tcp fairness for uplink and downlink flows in
  wlans,'' in {\em Global Telecommunications Conference, 2006. GLOBECOM'06.
  IEEE}, pp.~1--5, IEEE, 2006.

\bibitem{hiraguri2013}
T.~Hiraguri, K.~Nagata, T.~Ogawa, T.~Ueno, K.~Jin’no, and K.~Nishimori,
  ``Queuing scheme for improved downlink throughput on wlans,'' {\em Wireless
  Personal Communications}, pp.~1--16, 2013.

\bibitem{siwam2008}
S.~Siwamogsatham, ``Achieving uplink/downlink fairness in wlans via multiple
  backoff timers,'' in {\em Advanced Communication Technology, 2008. ICACT
  2008. 10th International Conference on}, vol.~3, pp.~1794--1797, IEEE, 2008.

\bibitem{virag10}
V.~Shah, N.~Mehta, and R.~Yim, ``Optimal timer based selection schemes,'' {\em
  Communications, IEEE Transactions on}, vol.~58, pp.~1814 --1823, june 2010.

\bibitem{wolf89}
R.~W. Wolff, {\em Stochastic Modeling and the Theory of Queues}.
\newblock New York, NY: Prentice-Hall, 1989.

\bibitem{wang2009joint}
L.-C. Wang, W.-C. Liu, A.~Chen, and K.-N. Yen, ``Joint rate and power
  adaptation for wireless local area networks in generalized nakagami fading
  channels,'' {\em Vehicular Technology, IEEE Transactions on}, vol.~58, no.~3,
  pp.~1375--1386, 2009.

\end{thebibliography}
\end{document}